\PassOptionsToPackage{table}{xcolor}

\documentclass[acmsmall,nonacm]{acmart}

\renewcommand\footnotetextcopyrightpermission[1]{} 
\usepackage{booktabs}
\usepackage{tabularx}
\usepackage{makecell}
\usepackage{enumitem}
\usepackage{threeparttable}
\usepackage{multirow}
\usepackage{colortbl}     
\usepackage{amsmath}
\usepackage{amsthm}
\usepackage{mathrsfs}
\usepackage[title]{appendix}
\usepackage{textcomp}
\usepackage{manyfoot}
\usepackage{algorithm}
\usepackage{algorithmicx}
\usepackage{algpseudocode}
\usepackage{listings}
\usepackage{longtable}
\usepackage{lscape}
\usepackage{array}
\usepackage{soul}
\usepackage{microtype}
\setcopyright{none}

\hypersetup{
  colorlinks=true,
  linkcolor=blue,
  citecolor=blue,
  urlcolor=blue
}

\usepackage[normalem]{ulem}  
\providecommand{\uline}[1]{\underline{#1}}

\usepackage{etoolbox}
\makeatletter

\AtBeginDocument{%
  \patchcmd{\@mkauthors@i}{\MakeUppercase}{\@firstofone}{}{}%
  \patchcmd{\@mkauthors@i}{\MakeUppercase}{\@firstofone}{}{}%

  \patchcmd{\@typeset@author@line}{\par\noindent}{\par\centering}{}{}%

  \patchcmd{\@mktitle@i}{\raggedright}{\centering}{}{}%
  \patchcmd{\@mktitle@i}{\noindent\@titlefont}{\@titlefont\centering}{}{}%
}

\makeatother

\begin{document}

\title{Comparative Analysis of Neural Retriever-Reranker Pipelines for Retrieval-Augmented Generation over Knowledge Graphs in E-commerce Applications}

\author{Teri Rumble}\email{teri.rumble@gmail.com}

\author{Zbyněk Gazdík}\email{gazdik.zbynek@gmail.com}
\author{Javad Zarrin}\authornote{Corresponding author}\email{j.zarrin@abertay.ac.uk}
\author{Jagdeep Ahluwalia}\email{j.ahluwalia@abertay.ac.uk}


\affiliation{%
  \institution{Abertay University}
  \streetaddress{Bell Street}
  \city{Dundee}
  \state{Scotland}
  \postcode{DD1 1HG}
  \country{UK}
}

\renewcommand{\shortauthors}{Rumble et al.}

\begin{abstract}
Recent advancements in Large Language Models (LLMs) have transformed Natural Language Processing (NLP), enabling complex information retrieval and generation tasks. Retrieval-Augmented Generation (RAG) has emerged as a key innovation, enhancing factual accuracy and contextual grounding by integrating external knowledge sources with generative models. Although RAG demonstrates strong performance on unstructured text, its application to structured knowledge graphs presents challenges: scaling retrieval across connected graphs and preserving contextual relationships during response generation. Cross-encoders refine retrieval precision, yet their integration with structured data remains underexplored. Addressing these challenges is crucial for developing domain-specific assistants that operate in production environments. This study presents the design and comparative evaluation of multiple Retriever-Reranker pipelines for knowledge graph natural language queries in e-Commerce contexts. Using the STaRK Semi-structured Knowledge Base (SKB), a production-scale e-Commerce dataset, we evaluate multiple RAG pipeline configurations optimized for language queries. Experimental results demonstrate substantial improvements over published benchmarks, achieving 20.4\% higher Hit@1 and 14.5\% higher Mean Reciprocal Rank (MRR). These findings establish a practical framework for integrating domain-specific SKBs into generative systems. Our contributions provide actionable insights for the deployment of production-ready RAG systems, with implications that extend beyond e-Commerce to other domains that require information retrieval from structured knowledge bases.

\end{abstract}


\maketitle

\section{Introduction}\label{sec1}
Natural Language Processing (NLP) has rapidly advanced with the use of Large Language Models (LLMs); however, conventional LLMs trained solely on static, publicly available data are prone to generating hallucinated or outdated content \cite{agrawal2024kghallucination}, especially when queries relate to proprietary or non‑public information \cite{ji2023survey}. Retrieval-Augmented Generation (RAG) addresses this limitation by augmenting LLMs with external knowledge retrieval, enabling the development of tools that integrate proprietary databases, non-public documentation, and domain-specific knowledge without requiring expensive model retraining or fine-tuning. This capability is particularly crucial for enterprise applications where a competitive advantage is based on the use of proprietary information (customer databases, product catalog or operational data) \cite{yang2024crag,lewis2020rag}. RAG transforms static LLMs into dynamic systems capable of grounding responses in authoritative, organization-specific knowledge while maintaining the fluency and reasoning capabilities of open-source LLMs. 

RAG works in two key stages: first, it dynamically retrieves a set of relevant documents based on the user query, and second, it integrates these documents into the input provided to the LLM, which generates a natural language response to the query \cite{lewis2020rag}. Although RAG bridges knowledge limitation problems, technical and operational limitations constrain practical deployment \cite{salemi2024urag,salemi2024erag}. These limitations are reflected in benchmarks such as the Comprehensive RAG Benchmark, which shows current state-of-the-art RAG systems achieve only 63\% accuracy, with straightforward, raw RAG implementations demonstrating 44\% accuracy \cite{yang2024crag}. These limitations become particularly pronounced when applying RAG to complex data structures, such as graphs with domain-specific knowledge representations \cite{edge2024graphrag}. Additionally, production-level systems require semantic reasoning capabilities, fast execution, and low computational costs to be scalable \cite{zhao2024ragdesign}.

The design of a RAG system needs to balance multiple trade-offs to optimize performance. Dense retrieval methods excel at capturing semantic similarity but often miss exact term matches, whereas sparse retrieval maintains high lexical precision, yet lacks semantic depth \cite{luan2021sparse}. At the synthesis stage, models are at risk of producing unsupported claims when integrating information from multiple sources \cite{agrawal2024kghallucination,zhang2025hallucination}. Finally, knowledge representation requirements vary widely across domains, creating technical barriers for general-purpose RAG systems, requiring domain-specific document management techniques \cite{aws2025_sparse_dense_rag}.

E‑commerce represents a particularly challenging domain for RAG implementation due to the scale, proprietary nature, and complex structural organization of product catalogs \cite{wu2024stark}. Commercial platforms often manage millions of products with constantly changing attributes such as pricing, product details, and customer reviews. Customer queries further increase complexity, ranging from precise product look-ups to vague requirement descriptions \cite{peng2024graphrag}. Context length constraints with large-scale tables often require document summarization or truncation, leading to information loss. High computational costs and long processing times associated with querying complex, relational data structures further complicate matters, as such operations often require expensive traversal and reasoning  procedures \cite{hu2024tkg,jiang2024diffkg,peng2024graphrag}.

The convergence of these limitations creates a need for specialized RAG architectures that can effectively handle semi-structured data. This study contributes to ongoing research in domain-specific information retrieval by conducting a comparative evaluation of several Retriever–Reranker architectures in the context of product information retrieval, using the Amazon STaRK Structured Knowledge Base (SKB)\footnote{The data used in this study are publicly available from the Hugging Face repository \texttt{snap-stanford/stark} at https://huggingface.co/datasets/snap-stanford/stark}, a representative proprietary dataset that captures detailed product attributes and inter-product relationships. Our primary contribution is the design and benchmarking of three Retriever–Reranker pipeline variants. The first, FRWSR (FAISS Retrieval and Webis Set-encoder Reranking \cite{schlatt2024setencoder}), is a dense retrieval pipeline that combines E5-Large embeddings, FAISS-HNSW retrieval, and Webis Set-Encoder/Large cross-encoder reranking. The second, FRMR (FAISS Retrieval and MS MARCO Reranking), is also a dense retrieval pipeline but instead integrates E5-Large embeddings, FAISS-HNSW retrieval, and MS MARCO MiniLM-L-6-v2 cross-encoder reranking. The third, BARMR (BM25 Augmented Retrieval and MS MARCO Reranking), adopts a sparse retrieval approach based on BM25, enhanced with edge node augmentation from the SKB graph structure to enrich the initial retrieval, followed by MS MARCO MiniLM-L-6-v2 cross-encoder reranking. The design choices are consistent with recent perspectives in the literature on LLM-based recommender systems and pipeline integration \cite{tang2025llmrec,lin2025llm4rec,chen2024offlinerl_tois}. 

We evaluate these models using a quantitative benchmarking framework against ground-truth node responses in the Amazon STaRK dataset, enabling direct comparison with published baselines \cite{wu2024stark}. The code associated with all three pipelines is available in a public Github repository \footnote{https://github.com/tep00018/ECommerceRAG}. Our contributions are threefold:
\begin{itemize}[left=2em]
    \item The design and implementation of multiple Retriever–Reranker pipelines that process natural language queries and retrieve responses exclusively from a proprietary SKB.
    \item An empirical performance evaluation on the STaRK benchmark, highlighting the comparative strengths and limitations of each pipeline to inform the development of scalable, enterprise-grade RAG systems.
    \item The introduction and technical specification of three pipeline configurations, FRWSR, FRMR, and BARMR, that integrate dense or sparse retrieval strategies with cross-encoder reranking.
\end{itemize}

The results offer practical insights into optimizing RAG models for leveraging SKB architectures in natural language query processing. The paper is organized as follows: an overview of RAG, its core components, and inherent limitations; a description of the Amazon STaRK dataset; a detailed methodology; the presentation of evaluation results; and a concluding discussion highlighting key findings and their practical implications.

\section{Background and Related Work}\label{sec2}
This section reviews the foundational concepts and related work necessary to contextualize the design and evaluation of the proposed Retriever–Reranker pipelines. This is explained in the context of retrieving information from a semi-structured knowledge graph, which is a key focus of this study.

\subsection{Retrieval Augmented Generation}
The original RAG framework proposed by Lewis et al. \cite{lewis2020rag} consists of two primary components: a Retriever and a Generator. Lewis' Retriever employs a dense retrieval strategy using a Dense Passage Retriever (DPR) \cite{karpukhin2020dense}, which encodes both the input query and candidate passages into a shared embedding space. Retrieval of top-ranked documents from an external corpus is based on vector similarity. The Generator, a pretrained sequence-to-sequence model such as BART, conditions its output on the retrieved documents. This enables the model to generate responses that are explicitly grounded in the external corpus, mitigating hallucination and improving relevance. Subsequent advancements in RAG systems have introduced more sophisticated retrieval pipelines, incorporating sparse retrieval (e.g., BM25) \cite{robertson2009probabilistic}, dense retrieval\cite{guu2020retrieval}, dense-sparse hybrids\cite{luan2021sparse}, uRAG (unified first-stage ranking for many RAGs), eRAG (document-level retrieval utility) and reranking mechanisms to enhance retrieval precision and output quality\cite{izacard2020leveraging}. 

Recent work further frames retrieval as a shared service for multiple RAG tasks and evaluates retrieval via document-conditioned downstream utility \cite{salemi2024urag,salemi2024erag}.
These improvements have expanded RAG’s applicability across a range of use cases, including domain-specific question answering, document summarization, and customer support automation \cite{hasan2025rag}.

\begin{figure}[htbp]
    \centering
    \includegraphics[width=\linewidth]{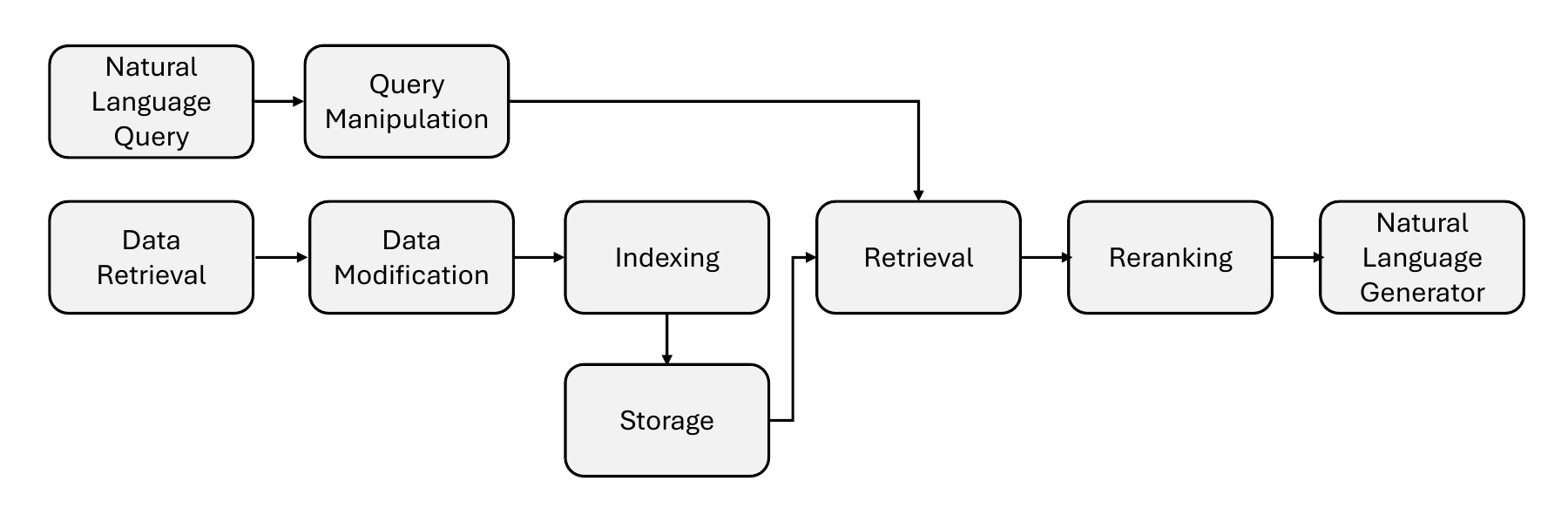}
    \caption{Example architecture of a Retrieval-Augmented Generation (RAG) system.}
    \Description{A schematic diagram showing how the retriever and generator components interact within a Retrieval-Augmented Generation system.}

    \label{fig:rag_architecure}
\end{figure}

Figure \ref{fig:rag_architecure} is a current view of a generic RAG pipeline. The system’s semi-structured knowledge source is subject to data modification operations (e.g., text cleaning, metadata enrichment) before being fed into an indexing component that constructs and maintains embedded representations in a vector database. When a user submits a natural language query, the query undergoes preprocessing steps such as tokenization, normalization, or query expansion, to optimize keyword and semantic matching. At query time, the manipulated query is compared against the indexed vector store to retrieve a candidate set of 'top k' entries via semantic (dense) or lexical (sparse) search. These candidates are then passed to a re-ranking module, which refines their order by computing relevance scores. Finally, the reordered results are provided to a natural language generator, which uses the selected knowledge snippets to produce a contextually relevant response \cite{lewis2020rag,yang2024crag}. 

While more accurate than LLMs alone, challenges remain when deploying RAG in more complex knowledge settings, such as large-scale relational knowledge graphs. These scenarios often involve multi-hop reasoning over structured entity relationships and edge metadata, requiring specialized retrieval strategies and discrete passage handling. Addressing these limitations continues to be an area of active research \cite{edge2024graphrag,wu2024stark}. To contextualize these challenges and illustrate current design strategies, Table \ref{tab:rag_sota} summarizes recent state‑of‑the‑art RAG architectures, outlining their pipeline components, distinguishing features, limitations, and representative application scenarios.

\begin{table*}[t]
\caption{RAG Pipeline State-of-the-Art Features and Limitations}
\label{tab:rag_sota}
\small
\begin{tabularx}{\textwidth}{@{}p{0.15\textwidth} p{0.26\textwidth} p{0.26\textwidth} p{0.27\textwidth}@{}}
\toprule
\textbf{Model} & \textbf{Pipeline Components} & \textbf{Key Features} & \textbf{Limitations} \\
\midrule

RAG \cite{lewis2020rag} & \raggedright DPR retriever, Dense vector index, Pretrained seq2seq BART generator & \raggedright Supports full sequence and per-token passage encoding & Dense-only retrieval struggles with structured/long-tail catalog data; DPR embeddings degrade on domain-specific entities. \\
\addlinespace

GraphRAG \cite{peng2024graphrag} & \raggedright Graph-based Indexing, Retrieval, Generation & \raggedright Leverages entity relations via graph neural methods to improve retrieval precision and context awareness & Graph construction and maintenance at catalog scale is expensive; inference over large graphs is too slow for production systems. \\

R\textsuperscript{2}AG \cite{ye2024r2ag} & \raggedright Standard retriever (BM25/DPR), LLM with retrieval-aware attention & \raggedright Retrieval-aware prompting strategy integrates retrieval information into LLM generation & LLM attention over large retrieved sets increases inference cost; accuracy depends heavily on retriever quality. \\
\addlinespace

IM-RAG \cite{yang2024imrag} & \raggedright Dense retriever (DPR), Sparse retriever (BM25), Multi-stage re-ranking, Generator & \raggedright Hybrid retrieval pipeline unifies sparse and dense signals for robust passage selection & Multi-stage re-ranking adds latency; lack of reranking step reduces accuracy. \\
\addlinespace

PlanRAG \cite{lee2024planrag} & \raggedright Retriever, Planner module, Generator & \raggedright Integrates explicit planning step to handle multi-step queries & Planning module introduces long inference chains and error propagation across steps. \\
\addlinespace

NIR-Prompt \cite{xu2023nirprompt} & \raggedright Multi-task IR framework with Essential Matching Module (EMM) and Matching Description Module (MDM) & \raggedright Decouples retrieval and reranking from generation to adapt across diverse IR tasks & High training cost for multi-task adaptation; inference slower due to multiple matching modules. \\
\addlinespace

Efficient Neural Ranking \cite{leonhardt2024efficient} & \raggedright Forward index retrieval, lightweight transformer encoder, re-ranking module & \raggedright Uses precomputed data representation with a compact encoder for low latency ranking & Optimized for speed but may sacrifice accuracy on nuanced queries; less effective for semantic similarity. \\

\bottomrule
\end{tabularx}
\end{table*}

\subsection{RAG Pipeline Components}
The retrieval mechanism in RAG systems operate by transforming natural language queries into vector representations and searching for similar entries within a pre-indexed vector database. These embeddings can be either sparse or dense, depending on the retrieval strategy employed. BM25 is a commonly used sparse retriever that ranks documents based on term frequency, inverse document frequency, and document length normalization. Sparse methods are computationally efficient, but they lack the ability to capture deeper semantic relationships between terms, relying on exact or near-exact keyword matches \cite{robertson2009probabilistic}. Dense retrieval utilizes neural networks to map tokens into low-dimensional continuous vectors, capturing semantic similarity and enabling the retriever to return relevant results even when the query and knowledge base wording differs \cite{karpukhin2020dense}. Dense methods excel in semantic generalization but are more resource-intensive in terms of training, storage, and retrieval computation. Beyond classical rerankers, setwise prompting with LLMs reduces token usage while retaining effectiveness \cite{zhuang2024setwise}.

Retrieval methods can be further enhanced by fusing retrieval with a re-ranking process in the pipeline. A re-ranking step prioritizes the information obtained using a traditional retrieval similarity search model, placing those that best answer the query in rank order. Re-ranking models are more powerful at classifying semantic similarity, but computationally too expensive and time-consuming to use solely for retrieval in large knowledge bases. This two-stage model leverages the speed of the retrieval models while optimizing the result with a more computationally expensive tool, such as an LLM, with semantic understanding capabilities \cite{karpukhin2020dense}. LLMs with an ability to reason over large volumes of text within a fixed context window are promising candidates for re-ranking tasks as they can be prompted to evaluate a given query in relation to a set of retrieved candidate documents and reorder them based on relevance \cite{zhao2025survey}.  However, using general-purpose LLMs for reranking can impractical in production due to context window limits and significant processing times and costs \cite{lin2025llmrecsys}.

Another class of re-ranking models includes sentence transformers, particularly bi-encoders and cross-encoders. While both are designed to model semantic similarity, bi-encoders enable efficient retrieval by independently encoding queries and documents, making them suitable for large-scale search tasks. In contrast, cross-encoders jointly encode query-document pairs, providing higher accuracy in relevance ranking, albeit at a significantly higher computational cost. \cite{wang2020minilm} \cite{bge_reranker_base}.

\subsection{Cross-encoder Comparison}
Cross-encoders serve as powerful reranking mechanisms in information retrieval pipelines. Unlike bi-encoders, which allow for previously computed document embeddings that can be stored in a vector database, cross-encoders compute relevance scores in real time by jointly encoding the query and document as a single input. This approach enables full self-attention across all tokens in both the query and the document at every layer of the transformer, allowing the model to capture fine-grained semantic interactions. The cross-encoder then outputs a single scalar relevance score for each query–document pair, making it particularly well-suited for re-ranking tasks. While more computationally intensive than bi-encoders, cross-encoders are still lighter and often more consistent than large language models (LLMs) when used specifically for re-ranking \cite{dejean2024reranking}. This makes them particularly effective for tasks requiring nuanced semantic understanding, including natural language inference (NLI) and passage ranking. Recent listwise/set-wise cross-encoders model inter-passage interactions efficiently, and intent-aware clarifying questions can further stabilize inputs. \cite{schlatt2024setencoder,zhao2024clarify}

The choice of cross-encoder type, point-wise, pair-wise, list-wise, or set-wise, determines the computational approach used during inference. Point-wise models process individual query-document pairs independently, requiring separate forward passes for each candidate document and optimizing relevance scores without consideration of inter-document relationships. Pair-wise approaches train models to distinguish relative preferences between document pairs for a given query, typically optimizing binary or margin-based loss functions that capture comparative relevance judgments \cite{wang2020minilm}. This makes a pair-wise approach computationally intensive due to the factorial nature of the calculation. List-wise models process entire ranked lists of candidate passages simultaneously, optimizing for position-aware metrics such as NDCG (Normalized Discounted Cumulative Gain) and MAP (Mean Average Precision) while capturing relative ordering relationships among multiple candidates through single forward passes. Pointwise cross-encoding methods are the most computationally efficient because they do not model interactions between candidates. Pair-wise and list-wise models are often infeasible for large texts due to memory input token constraints. \cite{ma2023zero}. 

\begin{table}[htbp]
\caption{Cross-encoder Types}
\label{tab:ce_types}
\small
\begin{tabularx}{\textwidth}{@{}p{0.12\textwidth} p{0.18\textwidth} p{0.22\textwidth} p{0.22\textwidth} p{0.12\textwidth}@{}}
\toprule
\textbf{Type} & \textbf{Input Construction} & \textbf{Attention Mechanism} & \textbf{Output Format} & \textbf{Primary Use} \\
\midrule
Point-wise & \raggedright Query + single document per forward pass & \raggedright Self-attention within query-doc pair & \raggedright Individual relevance score per document & Document reranking \\
\addlinespace
Pair-wise & \raggedright Query + two documents (triplet input) & \raggedright Joint attention over the query and two candidate documents & \raggedright Binary preference score ($A > B$) & Preference modeling \\
\addlinespace
List-wise & \raggedright Query + document list (sliding window) & \raggedright Sequential attention across ordered sequence & \raggedright Complete document ranking & Search result ranking \\
\addlinespace
Set-wise & \raggedright Query + all documents concatenated & \raggedright Inter-passage attention across collection & \raggedright Context-aware individual document scores & Document re-ranking \\
\bottomrule
\end{tabularx}
\end{table}

Set-wise architectures represent a new paradigm that operates over unordered collections of documents using permutation-invariant attention mechanisms to model inter-document relationships without relying on sequence position \cite{schlatt_2025}. See Table \ref{tab:ce_types} for comparison of cross-encoder types and See Table \ref{tab:model_overview} for an overview of state-of-the-art cross-encoders. Unlike point-wise models that require multiple independent forward passes, set-wise cross-encoders process all candidate documents within a single forward pass while employing inter-passage attention to generate context-aware individual scores for each document. This approach enables optimization for both relevance and diversity objectives, as each document's score reflects not only its individual query relevance but also its complementary value within the broader document set. 

\begin{table}[!htbp]
\centering
\caption{Overview of State-of-the-Art Cross-encoding Models}
\label{tab:model_overview}
\small
\begin{tabularx}{\textwidth}{@{}p{0.2\textwidth} p{0.10\textwidth} p{0.2\textwidth} p{0.38\textwidth}@{}}
\toprule
\textbf{Model} & \textbf{Type} & \textbf{Learning Objective} & \textbf{Architecture Notes} \\
\midrule
ms-marco-MiniLM-L-6-v2 & Point-wise & Passage Ranking & Distilled MiniLM with 6 layers trained on MS MARCO; optimized for speed and low latency. \\
\addlinespace
ms-marco-MiniLM-L-12-v2 & Point-wise & Passage Ranking & 12-layer version of above; better accuracy, used in sentence-transformers. \\
\addlinespace
amberoad/bert-multilingual-passage-reranking-msmarco & Point-wise & Passage Ranking & Multilingual BERT trained on MS MARCO; cross-lingual capable. \\
\addlinespace
BAAI/bge-reranker-base & Point-wise & Dense Retrieval amd Reranking & BERT-based reranker with list-wise contrastive learning trained on MS MARCO, Natural Questions; bilingual (EN+ZH). \\
\addlinespace
nli-deberta-v3-large & Point-wise Classification & Natural Language Inference (Entailment) & DeBERTaV3 with disentangled attention trained on MNLI, ANLI. \\
\addlinespace
nli-roberta-base & Point-wise Classification & Natural Language Inference (Entailment) & RoBERTa-base fine-tuned on SNLI, MNLI, FEVER-NLI datasets. \\
\addlinespace
stsb-roberta-large & Point-wise Regression & Semantic Textual Similarity & RoBERTa-large fine-tuned on STS Benchmark and SICK for cosine similarity on sentence pairs. \\
\addlinespace
webis/set-encoder-base & Set-wise & Novelty Detection & Permutation-invariant attention transformer encoder trained on Webis Novelty Corpus 2022; designed to assess novelty within text sets. \\
\addlinespace
webis/set-encoder-large & Set-wise & Novelty Detection & Larger version of set-encoder-base; better at modeling long-range dependencies. \\
\addlinespace
webis/set-encoder-novelty-base & Set-wise & Novelty Scoring & Set-encoder-base version tailored for document-level novelty detection in IR contexts. \\
\bottomrule
\end{tabularx}
\end{table}

\begin{figure}
    \centering
    \includegraphics[width=\textwidth]{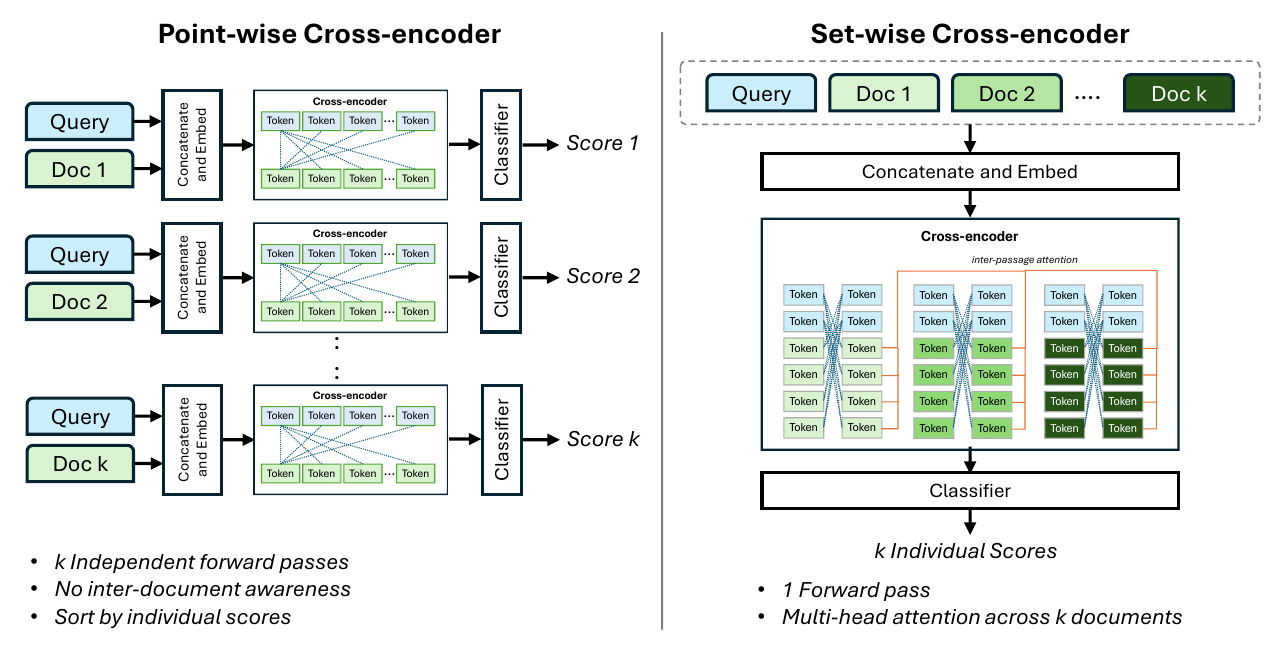}
    \caption{Point-wise and Set-wise cross-encoder comparison.}
    \label{fig:cross_encoder}
    \Description{Point-wise and Set-wise cross-encoder comparison.}
\end{figure}

These methodological distinctions significantly influence how models generalize across ranking tasks, with point-wise approaches emphasizing computational efficiency and set-wise methods prioritizing holistic optimization of result quality and coverage. See Figure \ref{fig:cross_encoder} for a comparison between point-wise and set-wise cross-encoding methods. In the context of product catalog reranking, the input context window is large, as each document comprises detailed product catalog information alongside associated customer reviews. Moreover, the necessity of retrieving and re-ranking multiple candidate documents imposes significant computational overhead, thereby precluding the use of pair-wise and list-wise cross-encoder architectures in this study. In practice, the best approach for a large knowledge base is either point‑wise or set‑wise reranking,—with point‑wise offering faster throughput, while set‑wise provides deeper context at the cost of longer processing times.

\subsection{Knowledge Graphs and Semi-Structured Knowledge}
When building a RAG pipeline, it is crucial to consider the structure and properties of the underlying knowledge source. In this study, knowledge retrieval is performed exclusively from a semi-structured knowledge base (SKB). We utilize the Amazon STaRK Semi-Structured Knowledge Base (SKB) and the associated STaRK Synthesized Prompt and Response Key datasets \cite{stark2024github, stark2024amazon}. These resources were selected due to their recency, scale, structural richness, and inclusion of a ground truth benchmark for natural language queries. The STaRK SKB serves as a representative example of a proprietary knowledge graph that would not be available to models trained solely on publicly accessible data. The SKB consists of over 1 million Node objects, each representing an Amazon product. These nodes encapsulate detailed metadata, including ASINs, titles, pricing, reviews, product features, and Q\&A content. In addition to this structured product information, the knowledge graph includes over 9 million directed edges that describe relationships such as "also-viewed" and "also-bought," allowing the representation of consumer browsing and purchase behaviors. This relational structure supports multi-hop reasoning, making it a strong test case for retrieval architectures that seek to go beyond flat document retrieval. The dataset also includes a set of test queries paired with ranked lists of correct node identifiers, enabling quantitative evaluation of retrieval accuracy. This explicit ground truth allows for model benchmarking without risk of data leakage and serves as an effective evaluation environment for Retriever and Reranker models. Complementary to retrieval over fixed SKBs, recent works edit or refine KGs to balance attributes and denoise relations in recommendation \cite{dou2024editkg,jiang2024diffkg}.

\section{Preliminaries}\label{sec3}
This section establishes notation (Table \ref{tab:notation}) and formal definitions for a formal RAG framework. RAG is decomposed into retriever, reranker and generator component. Evaluation metrics used to assess retrievers and rerankers are also defined.

\begin{table}[!ht]
\caption{Summary of Notation: We adopt standard IR/RAG notation and primitives including queries and corpora as in classic IR \citep{salton1975vector,manning2008introduction}, retrievers \(R\) instantiated as BM25 or dense DPR \citep{robertson2009probabilistic,karpukhin2020dense}, top-\(k\) selection \citep{fagin2001optimal}, knowledge graphs and entity linking \citep{auer2007dbpedia,bollacker2008freebase,hoffart2011robust}, re-ranking with learning-to-rank and cross-encoders \citep{burges2005learning,nogueira2019passage,cao2007learning,xia2008listwise}, generation via retrieval-augmented language models \citep{lewis2020rag,guu2020realm}, and evaluation with Hit@\(k\), Recall@\(k\), and MRR \citep{he2017neural,manning2008introduction,voorhees2000trec8}.
}\label{tab:notation}
\centering
\small
\begin{tabular}{@{}p{0.47\textwidth} p{0.49\textwidth}@{}}
\begin{minipage}[t]{\linewidth}\vspace{0pt}
\raggedright
\begin{tabular}{@{}>{\raggedright\arraybackslash}p{0.28\linewidth}@{}>{\raggedright\arraybackslash}p{0.68\linewidth}@{}}
\toprule
\textbf{Notation} & \textbf{Explanation} \\
\midrule
$q \in \mathcal{Q}$ & Query (from distribution $q \!\sim\! P_q$) \\[4pt]
$\mathcal{D}$ & Document corpus (unstructured text) \\[4pt]
$a \in \mathcal{A}$ & Generated answer / model output \\[4pt]
$R$ & Retriever mapping $q \mapsto C_k(q)$ \\[4pt]
$C_k(q)$ & Top-$k$ retrieved candidate set ($|C_k(q)|=k$) \\[4pt]
$d^+, d^-$ & Relevant and non-relevant documents (w.r.t.\ a query $q$) \\[4pt]
$\mathcal{G}=(\mathcal{V},\mathcal{E})$ & Knowledge graph with vertices $\mathcal{V}$ and edges $\mathcal{E}$ \\[4pt]
$\eta(\cdot)$ & Document-to-vertex mapping (entity linking) \\[4pt]
$\mathcal{N}_{\mathcal{G}}^{(1)}(v)$ & 1-hop neighbor set of node $v$ in $\mathcal{G}$ \\[4pt]
$\mathrm{Expand}_{\mathcal{G},1}(S)$ & 1-hop expansion operator on set $S$ (deduplicated) \\[4pt]
\bottomrule
\end{tabular}
\end{minipage}
&
\begin{minipage}[t]{\linewidth}\vspace{0pt}
\raggedright
\begin{tabular}{@{}>{\raggedright\arraybackslash}p{0.30\linewidth}@{}>{\raggedright\arraybackslash}p{0.66\linewidth}@{}}
\toprule
\textbf{Notation} & \textbf{Explanation} \\
\midrule
$C_k^{\mathcal{G}}(q)$ & KG-augmented candidate set after expansion \\[4pt]
$f(q,d)$ & Reranker scoring function \\[4pt]
$\pi_f$ & Permutation on $C_k(q)$ induced by $f$ \\[4pt]
$d_j$ & Document at rank $j$ in reranked order \\[4pt]
$G$ & Generator: autoregressive LLM conditioned on $q$ and context \\[4pt]
$y(q,d)$ & Relevance label of document $d$ for query $q$ \\[4pt]
$\mathrm{Hit@}k$ & Indicator that a relevant doc appears in top-$k$ after reranking \\[4pt]
$\mathrm{Recall@}k$ & Probability that a relevant doc is included in top-$k$ retrieved set \\[4pt]
$\mathrm{MRR}$ & Mean Reciprocal Rank of the first relevant document \\[4pt]
\bottomrule
\end{tabular}
\end{minipage}
\end{tabular}
\end{table}

\subsection{RAG Framework}
Formally, given a query $q \in \mathcal{Q}$ and a large corpus $\mathcal{D}$, the goal is to generate an output $a \in \mathcal{A}$ that is factually consistent with respect to $\mathcal{D}$. Unlike pure language modeling, RAG explicitly decomposes the problem into retrieval, reranking, and generation, reducing hallucination.

\paragraph{Retriever.}
A retriever $R$ maps the query $q$ to a candidate set of $k$ documents:
\begin{equation}\label{eq:retriever}
C_k(q) \;=\; R(q) \;\subset\; \mathcal{D}, \qquad |C_k(q)| = k.
\end{equation}

\emph{Optional KG expansion.}  
The retriever produces an initial candidate set $C_k(q)$, which may be used directly in downstream reranking and generation. In SKBs, retrieved documents correspond (via entity linking) to vertices in a knowledge graph $\mathcal{G}=(\mathcal{V},\mathcal{E})$. To leverage structured relations, we may expand $C_k(q)$ by appending 1-hop neighbors from $\mathcal{G}$. Let $\mathcal{N}_{\mathcal{G}}^{(1)}(v)=\{u\in\mathcal{V}:(v,u)\in\mathcal{E}\}$ denote the 1-hop neighbor set of node $v\in\mathcal{V}$, and let $\mathrm{Expand}_{\mathcal{G},1}(\cdot)$ be the expansion operator. The augmented set is
\begin{equation}\label{eq:retriever_aug}
C_k^{\mathcal{G}}(q) \;=\; R_{\mathcal{G}}(q)
\;=\;
\mathrm{Expand}_{\mathcal{G},1}\big(C_k(q)\big)
\;=\;
C_k(q)\ \cup\ \bigcup_{v\in \eta(C_k(q))} \mathcal{N}_{\mathcal{G}}^{(1)}(v),
\end{equation}

where \(\eta(\cdot)\) denotes the document-to-vertex mapping (entity linking). Since unions are over sets, duplicates are removed, ensuring $C_k^{\mathcal{G}}(q)$ is a set of unique nodes/documents.

The KG-augmented retriever $R_{\mathcal{G}}$ therefore extends conventional retrieval with an \emph{optional} structured expansion step, providing semantically connected candidates that rerankers and generators can exploit for improved factual grounding and entity disambiguation.

\paragraph{Reranker.} A reranker $f$ assigns fine-grained scores to candidates $d \in C_k(q)$, inducing a permutation $\pi_f$:

\begin{equation}\label{eq:reranker}
\begin{aligned}
\pi_f &: C_k(q)\to\{1,\dots,k\}, \\
d_j &\text{ denotes the document with } \pi_f(d_j)=j,
\text{ so the ordered list is } (d_1,\dots,d_k).
\end{aligned}
\end{equation}

Cross-encodera are often implemented as rerankers, jointly encoding $(q,d)$ pairs. 

\paragraph{Generator.}
Finally, a generator $G$ conditions on the query $q$ and the reranked set $C_k(q)$ to produce an answer:

\begin{equation}\label{eq:generator}
a = G\!\left(q, \pi_f(C_k(q))\right).
\end{equation}

The generator is typically an autoregressive LLM, trained with maximum likelihood estimation (MLE) but often fine-tuned with reinforcement learning from human feedback (RLHF) or retrieval-augmented objectives to balance fluency and factuality.

\subsection{Evaluation Metrics}
To evaluate the effectiveness of retrievers and rerankers in a RAG pipeline, we employ several complementary measures. These metrics capture both the retriever's ability to surface relevant documents and the reranker's ability to correctly prioritize them. A summary description of each metric is given in Table \ref{tab:eval_metrics_2} and in Equations \ref{eq:hit1_eg}-\ref{eq:mrr_expectation}.

\begin{table}[h]
\caption{Evaluation Metrics}
\label{tab:eval_metrics_2}
\small
\begin{tabularx}{\textwidth}
{@{}lp{0.8\linewidth}@{}}
\toprule
\textbf{Metric} & \textbf{Description} \\
\midrule
Hit @ 1 & Assesses whether the model’s top prediction result is in the answer key to evaluate the relevance of the top recommendation. \\
\addlinespace
Hit @ 5 & Assesses whether any of the model’s top 5 prediction results are in the answer key to evaluate the top 5 recommendation relevance. \\
\addlinespace
Recall @ 20 & Measures the proportion of relevant items in the top 20 prediction results, measuring the model’s ability to identify all relevant items. \\
\addlinespace
MRR & Calculates the reciprocal of the rank at which the first relevant item appears in the list of predictions, emphasizing the rank of the first correct answer. \\
\bottomrule
\end{tabularx}
\end{table}

\paragraph{Hit@k.}
Hit@k evaluates whether a relevant document appears within the top-$k$ positions of the reranked list:

\begin{equation}\label{eq:hit1_eg}
\mathrm{Hit@}k\big(f \mid q, C_k\big)
\;=\;
\mathbf{1}\!\left\{ \exists\, d \in \{d_1,\dots,d_k\} : y(q,d)=1 \right\},
\end{equation}
where $(d_1,\dots,d_k)$ denotes the reranked ordering induced by $f$.


\paragraph{Recall@k.}
The recall at cutoff $k$ measures the probability that at least one relevant document is included in the top-$k$ retrieved set:

\begin{equation}\label{eq:recall_per_query}
\mathrm{Recall@}k\big(f\mid q,C\big)
\;=\;
\frac{\displaystyle\sum_{j=1}^{k} y(q,d_j)}{\displaystyle R_q},
\quad\text{where }R_q:=\sum_{d\in C} y(q,d)
\end{equation}

\paragraph{Mean Reciprocal Rank (MRR).}
MRR captures the average inverse rank of the first relevant document across queries:

\begin{equation}\label{eq:mrr_expectation}
\mathrm{MRR}(f)
\;=\;
\mathbb{E}_{q}\!\left[ \frac{1}{\min\{\,j \ge 1 : y(q,d_j) = 1\,\}} \right],
\end{equation}

where $d_j$ is the $j$-th document in the reranked list. This metric rewards systems that place relevant documents as high as possible.

\medskip
Hit@k, Recall@k, and MRR provide a comprehensive view of retrieval and reranking performance: Recall@k emphasizes coverage, Hit@k emphasizes top-$k$ placement, and MRR emphasizes ranking quality. 

\section{Experimental Methodology}
Evaluation of retrieval pipelines requires both a high-quality application representative dataset and a benchmarking framework. The Amazon SKB and its companion Synthesized Prompt and Response Key \cite{wu2024stark} offer a unique testbed by combining rich product metadata, inter-product relationships, and query–answer mappings. Using this established resource, we ensure comparability with previous work.

This study employs a systematic three-phase experimental methodology to identify and validate optimal Retriever-Reranker pipelines for knowledge extraction from the Amazon STaRK Semi-Structured Knowledge Base. The research framework is designed to comprehensively evaluate both individual component performance and integrated pipeline effectiveness across diverse retrieval scenarios. The experimental design follows a sequential optimization approach comprising three distinct phases:
\begin{itemize}[left=2em] 
    
    \item Phase 1: Retrieval Model Selection and Comparison - involves the selection and initial assessment of candidate retrieval models. This phase establishes the foundation for subsequent experimentation by identifying promising retriever models based on theoretical suitability and computational feasibility.
    \\ 
    \item Phase 2: Re-ranking Model Selection and Comparison - focuses on rigorous benchmarking of selected re-ranking models using the validation dataset (n=910 queries). This controlled evaluation phase enables model selection while preventing overfitting to the final evaluation sets. Performance metrics from this phase inform the selection of the highest-performing re-ranking model for integration into complete retrieval pipelines.
    \\ 
    \item Phase 3: Integrated Pipeline Evaluation - constructs and evaluates complete end-to-end retrieval systems by combining the selected retrieval method in Phase 1 (with and without KG expansion) with the re-ranking models identified in Phase 2. These validated pipeline configurations undergo a comprehensive performance assessment using both evaluation query sets to ensure robust comparison with existing benchmarks and thorough characterization of system capabilities.
\end{itemize}

This methodological framework ensures systematic progression from component-level optimization to integrated system evaluation, providing both rigorous model selection criteria and comprehensive performance validation across multiple evaluation contexts.

\subsection{Dataset Selection}
 This analysis utilized two primary datasets from the STaRK (Semi-structured Table-to-text Retrieval and Question-answering Knowledge) framework: the Amazon Semi-Structured Knowledge Base (SKB) \cite{stark2024github} and the STaRK Synthesized Prompt and Response Key \cite{stark2024amazon}, both accessible through the snap-stanford/stark repository on Hugging Face \cite{snap2024huggingface}.
 
 The SKB represents a comprehensive e-Commerce knowledge graph comprising nearly 1M nodes of information. See Table \ref{tab:skb_node_attributes} for an example of a product node. The dataset is predominantly composed of product nodes, each containing structured information including product titles, pricing data, and brand associations. There are four distinct entity types (product, brand, color, and category) interconnected through five relationship types (Table \ref{tab:node_types}). Specifically, the graph includes behavioral relationships ('also-bought' and 'also-viewed') between product entities, as well as categorical relationships ('has-brand', 'has-color', and 'has-category') linking products to their respective attributes. 

\begin{table}[htbp]
\caption{SKB Node Attributes}
\label{tab:skb_node_attributes}
\centering
\small
\begin{tabularx}{\textwidth}{@{}lllp{0.4\textwidth}@{}}
\toprule
\textbf{Column Name} & \textbf{Non-null Count} & \textbf{Data Type} & \textbf{Example} \\
\midrule
node\_id & 1,035,542 & Integer & 864782 \\
node\_type & 1,035,542 & Integer & 0 \\
asin & 957,192 & String & B003R8K6SS \\
title & 957,192 & String & Rhode Gear Super Shuttle 2 Bike Rack \\
global\_category & 957,192 & String & Sports and Outdoors \\
category & 727,527 & List & ['bike racks'] \\
price & 957,192 & String & \textit{blank} \\
brand & 860,361 & String & Rhode Gear \\
color & 174,068 & List & None \\
feature & 957,192 & List & ['Securely carries two bicycles', 'Fits 2 inch and 1.25 inch trunks...'] \\
rank & 880,325 & String & 950,041 in Sports \& Outdoors \\
details & 957,192 & String & \{"product\_dimensions": "22 x 22 x 7 inches ; 14 pounds", "shipping\_weight": "14.5 pounds"...\} \\
description & 957,192 & List & ['An affordable and secure solution for carrying two bikes...'] \\
review & 957,192 & List & [\{'reviewerID': 'A2B157KE7AYTO', 'summary': 'very solid'...\}] \\
qa & 957,192 & List & [\{'questionType': 'yes/no', 'question': 'Can this be attached to a truck?'...\}] \\
also\_buy & -- & List & [955843, 955853, 955832, 955834, 955835, 955836, 955837, 955838] \\
also\_view & -- & List & [] \\
\bottomrule
\end{tabularx}
\end{table}

\begin{table}[htbp]
\caption{Node Types}
\label{tab:node_types}
\centering
\small
\begin{tabularx}{\textwidth}{@{}lp{0.7\textwidth}@{}}
\toprule
\textbf{Node Type} & \textbf{Examples} \\
\midrule
0: 'product' & See Table~\ref{tab:skb_node_attributes} for Node Example: 864782 \\
\addlinespace
1: 'brand' & Node: 1029490: \{"brand\_name": "bdd"\} \\
& Node: 1011310: \{"brand\_name": "SUNSECT"\} \\
& Node: 959573: \{"brand\_name": "Advanced Fire Starter Pro"\} \\
\addlinespace
2: 'category' & Node: 1033665: \{"category\_name": "tactical vests"\} \\
& Node: 1033808: \{"category\_name": "whistles"\} \\
& Node: 1033600: \{"category\_name": "standard skateboards"\} \\
\addlinespace
3: 'color' & Node: 1034392: \{"color\_name": "deep indigo"\} \\
& Node: 1034100: \{"color\_name": "blackberry"\} \\
& Node: 1035457: \{"color\_name": "washington capitals"\} \\
\bottomrule
\end{tabularx}
\end{table}

The STaRK Synthesized Prompt and Response Key comprises 9,100 natural language queries paired with their corresponding product nodes, ranked in order of relevance to each query. Representative examples are shown in Table~\ref{tab:stark_prompt_dataset}. This query–response format enables consistent and comparable benchmarking across different pipeline variants. Importantly, the evaluation protocol involves no generative components; performance is assessed exclusively on the retrieval–reranker pipeline prior to any generation, using Hit@k, Recall@k, and MRR as the primary measures.

\begin{table}[htbp]
\caption{Amazon STaRK Synthesized Prompt and Ground Truth Dataset}
\label{tab:stark_prompt_dataset}
\centering
\small
\begin{tabularx}{\textwidth}{@{}p{0.6\textwidth}p{0.3\textwidth}@{}}
\toprule
\textbf{Query} & \textbf{Answer Node IDs} \\
\midrule
What are some high-quality, pocket-sized map sets with detailed features ideal for outdoor adventures? & [258, 943747, 68, 101, 629637, 102, 38406, 629620, 629624, 629625, 271900, 271901] \\
\addlinespace
Is there a set of Colorado 14ers trail maps that include both UTM grid and latitude-longitude ticks available? & [130947, 771637, 102] \\
\addlinespace
What are some visually stunning castle card modelling kits produced by Aue Verlag? & [104] \\
\addlinespace
Can you suggest a high-quality OEM Control unicycle featuring a powder-coated finish steel fork that is also easy to store in a corner? & [112, 268935, 268925, 268927] \\
\bottomrule
\end{tabularx}
\end{table}

\subsection{Evaluation Benchmarks}
Benchmarks that emulate proprietary or semi-structured knowledge repositories are scarce, yet they are vital for rigorously assessing Retrieval-Augmented Generation (RAG) pipelines in realistic settings. The Amazon STaRK Semi-Structured Knowledge Base (SKB) and its Synthesized Prompt and Response Key dataset \cite{wu2024stark} provide a unique benchmark by combining comprehensive product metadata, explicit inter-product relationships, and a predefined ground truth mapping of natural language queries to ranked node lists. Wu et al. \cite{wu2024stark} reported baseline performance on the STaRK synthetic query set using various retrieval and RAG methods. We adopt their evaluation metrics  (Table \ref{tab:eval_metrics_2}) to enable direct comparisons.

To ensure methodological consistency with established benchmarks, we adopted the evaluation protocol established by Wu et al., which excludes queries containing more than 20 ground truth answers. This filtering criterion addresses a known bias in retrieval evaluation metrics, where queries with extensive ground truth sets can artificially inflate Recall@20 performance while simultaneously suppressing Hit Rate and Mean Reciprocal Rank (MRR) scores due to the increased likelihood of relevant documents appearing beyond the top-k positions.

Following this protocol, we systematically partitioned the complete 9,100-query dataset into three distinct evaluation subsets to support comprehensive performance assessment:

\begin{itemize}
    \item \textbf{Validation Set} (n=910): A stratified 10\% sample of queries with answer lengths $\leq 20$, reserved for hyperparameter tuning and intermediate model benchmarking.
    \item \textbf{Evaluation Set 1} (n=6,380): The complete filtered dataset containing all queries with answer lengths $\leq 20$, used for primary performance evaluation and direct comparison with Wu et al.'s results.
    \item \textbf{Evaluation Set 2} (n=9,100): The unfiltered complete dataset including all queries regardless of ground truth set size, employed to assess model robustness across varying query complexity levels.
\end{itemize}

This tri-partite evaluation framework enables both rigorous comparison with existing benchmarks and comprehensive assessment of model performance across diverse query characteristics. All queries are posed in natural language and paired with ordered node ID lists as ground truth. Table \ref{tab:baseline_results} reproduces the published baseline results specific to a random 10\% sample of the the Amazon STaRK SKB \cite{wu2024stark}, against which our proposed Retriever-Reranker pipelines are compared.

\begin{table}[h!]
\caption{Baseline RAG Results on Amazon STaRK Synthetic Query Set \cite{wu2024stark}}
\label{tab:baseline_results}
\small
\centering
\begin{tabularx}{\textwidth}{@{}p{0.22\textwidth} p{0.22\textwidth} 
                  p{0.11\textwidth} p{0.11\textwidth} p{0.11\textwidth} p{0.11\textwidth}@{}}
\toprule
\textbf{Model} & \textbf{Type} & \textbf{Hit@1} & \textbf{Hit@5} & \textbf{Recall@20} & \textbf{MRR} \\
\midrule
BM25 & Retrieval & 42.68 & 67.07 & 54.48 & 54.02 \\
DPR (roberta) & Retrieval & 16.46 & 50.00 & 42.15 & 30.20 \\
ANCE (roberta) & Retrieval & 30.09 & 49.27 & 41.91 & 39.30 \\
QAGNN (roberta) & Retrieval & 25.00 & 48.17 & 51.65 & 36.87 \\
ada-002 & Retrieval & 39.02 & 64.02 & 49.30 & 50.32 \\
voyage-12-instruct & Retrieval + Re-ranking & 43.29 & 67.68 & \uline{56.04} & 54.20 \\
LLM2Vec & Retrieval & 18.90 & 37.80 & 34.73 & 28.76 \\
GritLM-7b & Retrieval + Reranking & 43.29 & \textbf{71.34} & \textbf{56.14} & 55.07 \\
multi-ada-002 & Retrieval & 40.85 & 62.80 & 52.74 & 51.54 \\
ColBERTv2 & Retrieval & 44.31 & 65.24 & 51.00 & 55.07 \\
Claude3 Reranker & Retrieval + Reranking & \textbf{45.49} & 71.13 & 53.77 & \textbf{55.91} \\
GPT4 Reranker & Retrieval + Reranking & \uline{44.79} & \uline{71.17} & 55.35 & \uline{55.69} \\
\bottomrule
\end{tabularx}
\end{table}

\begin{figure}[htbp]
    \centering
    \includegraphics[width=\linewidth]{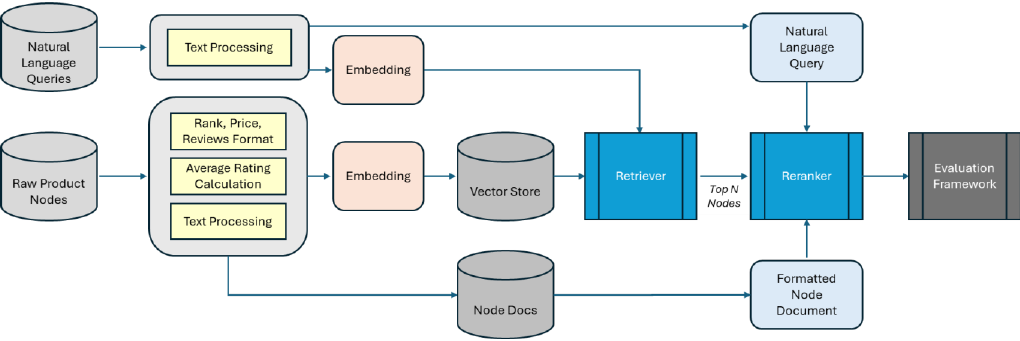}
    \caption{Model Evaluation Pipeline Architecture.}
    \label{fig:RAG_model_eval_pipeline}
    \Description{Model Evaluation Pipeline Architecture.}
\end{figure}

Each model reads ‘query’ questions from a separate file, retrieves relevant information from the knowledge vector database, generates answers using the retrieval model, re-ranks the nodes based on similarity with the query, and logs the output for both the Retriever and the Reranker in the form of product nodes. This is then measured against the gold standard answer set. A high-level architecture of the pipeline structure used in this evaluation is illustrated in Figure \ref{fig:RAG_model_eval_pipeline}. 

\subsection{Data Preprocessing}
For computational processing and retrieval optimization, the graph-structured data was transformed into a tabular DataFrame format, with each entity attribute represented as a distinct column. To ensure consistency across the retrieval pipeline, node identifiers were preserved as locators that maintain alignment between the structured data, corresponding vector embeddings, and documents used in subsequent re-ranking procedures. The pre-processing pipeline was designed to optimize textual content for retrieval effectiveness. Initial experiments using the BM25 algorithm were conducted to evaluate the impact of various node field combinations and text pre-processing strategies on retrieval performance, informing the selection of optimal pre-processing parameters for the full experimental framework.

The BM25 model and sparse embeddings were used to debug and test the efficacy of incorporating specific node fields and text preprocessing options (Table \ref{tab:preprocessing_methods}). 

\begin{table}[htbp]
\caption{Data Processing Methods}
\label{tab:preprocessing_methods}
\small
\begin{tabularx}{\textwidth}
{@{}p{4cm}p{9.5cm}@{}}
\toprule
\textbf{Process} & \textbf{Steps} \\
\midrule
Text Processing & For each text field, application of a combination of methods: lowercase, special characters removed, contractions expanded to plain text, punctuation removed, stop words removed, digits and digit words removed, lemmatization. \\
\addlinespace
Rank Formatting & Strip text and punctuation. \\
\addlinespace
Price Formatting & Remove \$ to represent as a float. \\
\addlinespace
Review Formatting & Extract and append written text for each reviewer. \\
\addlinespace
Average Rating Calculation & Calculate the mean rating for each product node. \\
\addlinespace
Node Edge Properties & Log  ‘also\_buy’, ‘also\_view' associations. \\
\addlinespace
Document Creation & Simple concatenation of selected fields pre-appended with the field label ("Brand:", "Title:" "Features:", etc.). \\
\bottomrule
\end{tabularx}
\end{table}

\begin{figure}[htbp]
    \centering
    \includegraphics[width=1\linewidth]{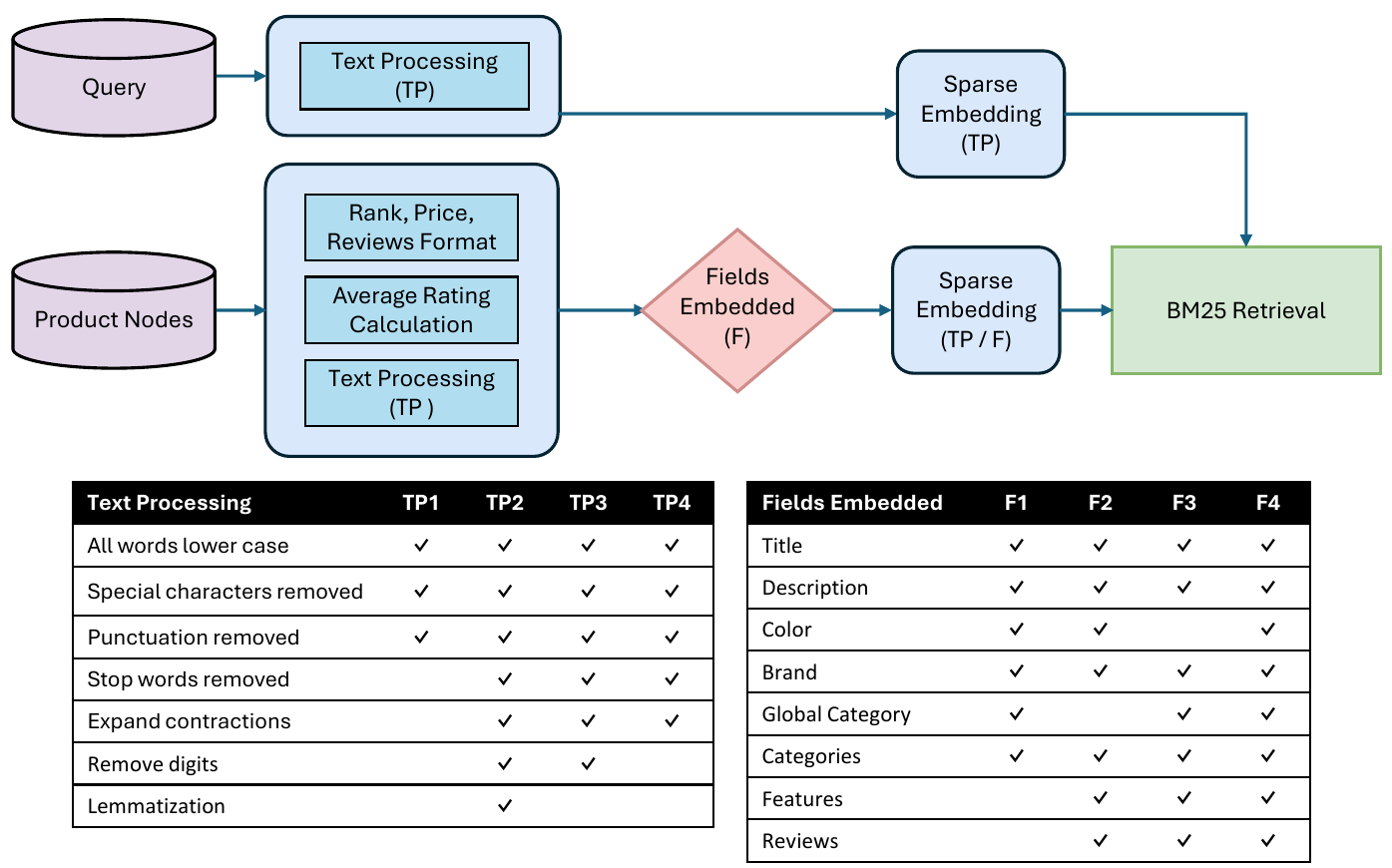}
    \caption{BM25 Sparse Embedding Pipeline Optimization}
    \label{fig:text_processing}
    \Description{BM25 Sparse Embedding Pipeline Optimization}
\end{figure}

We began by preparing the Amazon STaRK semistructured knowledge base (SKB) through text preprocessing using various options. These options were evaluated by utilizing a sparse embedding vector store and BM25 retrieval. Both sparse (BM25) and dense (vector embedding) indices were then constructed to support retrieval. For the dense index, we used the E5-Large embedding model. We next evaluated each retrieval model by tuning key hyperparameters on a validation set. The top performing retriever was subsequently paired with candidate rerankers (cross-encoders and LLMs). Finally, the best reranker was selected and four retriever/reranker pipelines were benchmarked against both the original and modified STaRK test sets using standard metrics (Hit@k, Recall@k, MRR) and compared to published baselines.

\begin{figure}[htbp]
    \centering
    \includegraphics[width=1\linewidth]{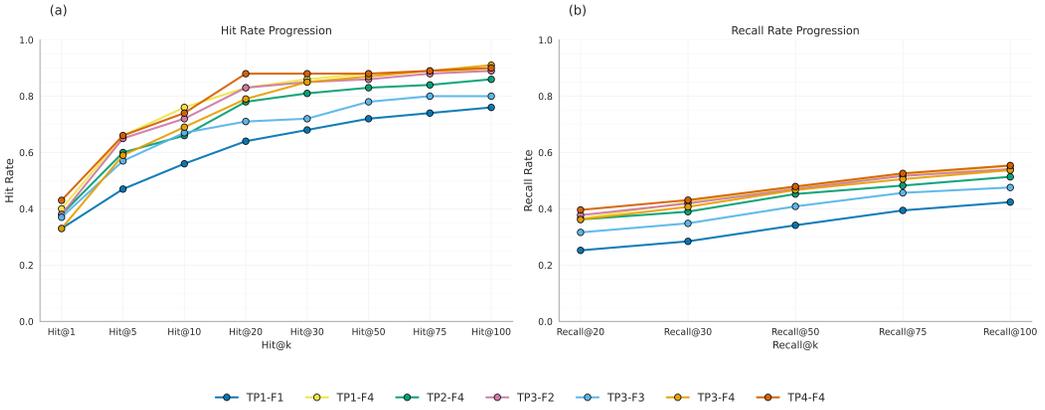}
    \caption{Text Pre-processing Method Comparison with 100 queries.}
    \label{fig:text_preprocessing_comparison}
    \Description{Text Pre-processing Method Comparison with 100 queries.}
\end{figure}

The BM25 retrieval method was used to assess text pre-processing methods and optimal field inclusion to optimize the SKB into a ‘clean’ tabular format. A comparison of four different pre-processing methods utilizing four field combinations was evaluated (Figure \ref{fig:text_preprocessing_comparison}). 

Experimental results (Figure \ref{fig:text_preprocessing_comparison}) demonstrate a clear advantage to using all the text fields in the SKB, especially the customer reviews (method F4). Lemmatization and removing digits did not improve results. Therefore, method TP4 was used for text processing going forward.

\subsection{FAISS HNSW Hyperparameter Tuning}
Two HNSW indexes were constructed for evaluation. This indexing method builds a graph using the following parameters: M: Controls the number of neighbors each node can have in the graph, efConstruction: Number of candidate neighbors considered during index construction, efSearch: Number of candidate neighbors explored during search. Evaluation metrics for a Flat index (brute force) and two HNSW indexes are in Table \ref{tab:faiss_results}.

\begin{table}[h!]
\caption{FAISS Index Comparison Results (*910 Query Set)}
\label{tab:faiss_results}
\small
\begin{tabularx}{\textwidth}{@{}lrrrrrrr@{}}
\toprule
\textbf{Index} & \textbf{M} & \textbf{efConstruction} & \textbf{efSearch} & \textbf{Hit@1} & \textbf{Hit@5} & \textbf{Recall@20} & \textbf{MRR} \\
\midrule
FAISS FLAT & N/A & N/A & N/A & \textbf{0.4363} & \textbf{0.7088} & \uline{0.4148} & \textbf{0.5587} \\
FAISS HNSW & 32 & 40 & 16 & 0.2923 & 0.5011 & 0.4146 & 0.3890 \\
FAISS HNSW & 64 & 100 & 200 & \uline{0.3879} & \uline{0.5385} & \textbf{0.5351} & \uline{0.5060} \\
\bottomrule
\end{tabularx}%
\end{table}

\section{Results}\label{sec_results}
This section presents the experimental results of our evaluation of retrieval and reranking methods for the Amazon STaRK SKB. Our analysis proceeds in three stages, each building upon the previous to construct an optimized retrieval-augmented generation pipeline. First,  we assessed multiple retrieval methods, including sparse lexical approaches (BM25 with and without graph-based augmentation) and dense vector methods (FAISS with exact and approximate search). Performance was measured using standard benchmark metrics on the 910 query validation set. Second, the re-ranking methods, comparing the effectiveness of various cross-encoder architectures and an LLM. Finally, the best retriever-reranker pipelines are compared on the complete 9,100 query evaluation dataset. Throughout these analyses, we present comparative performance, statistical significance tests, and implications for pipeline optimization.

\subsection{Retrieval Model Comparison}
Wu et al. tested a range of retrieval and LLM-based re-ranking methods on this database with mixed results \cite{wu2024stark}. While the methods they used generally performed well in retrieving entity data, the authors identified a limitation in the sufficient processing of edge data. The present project, therefore, aims to design a retrieval pipeline that can effectively search the Amazon STARK SKB, which will involve testing, comparing, and combining several retrieval methods ranging from simple bag-of-words keyword matching to sophisticated dense embeddings that represent semantic similarity retrieval methods. Priority considerations for selecting a Retrieval model: 1) applicability to semi-structured text data, 2) ability to apply semantic understanding, and 3) processing efficiency. BM25 and FAISS methods were selected for further experimentation (See Table \ref{tab:retrieval_models}).

To further augment BM25, a 1-hop expansion of nodes retrieved is tested to determine with 'also buy' and 'also view' node relations enable better retrieval. FAISS FLAT and HNSW indexing methods were selected. The Hierarchical Navigable Small World (HNSW) algorithm utilizes a multi-layer graph structure for efficient approximate nearest neighbor estimation. The FAISS FLAT algorithm performs a brute-force nearest neighbor search by computing distances between the query vector and all vectors in the index.

\renewcommand{\arraystretch}{1.2} 

\begin{table}[htbp]
\begin{threeparttable}
\small
\caption{Retrieval Models Evaluated \cite{robertson2009probabilistic, faiss2025welcome}}
\label{tab:retrieval_models}
\centering
\begin{tabularx}{\textwidth}{@{}p{0.15\textwidth} p{0.2\textwidth} p{0.28\textwidth} p{0.28\textwidth}@{}}
\toprule
\textbf{Model} & \textbf{Type} & \textbf{Strengths} & \textbf{Weaknesses} \\
\midrule
BM25 & \raggedright Lexical (sparse retrieval) & \raggedright Efficient and interpretable; requires exact keyword matching. & Limited to exact term matching; lacks capture of semantic similarity between terms. \\
\addlinespace
BM25 Augment\tnote{a} & \raggedright Sparse lexical retrieval plus 1-hop knowledge graph expansion & \raggedright BM25 with 1-hop relational expansion to augment lexical matches. & Increased computational overhead due to expanded candidate retrieval. \\
\addlinespace
FAISS - FLAT & \raggedright Dense retrieval (vector-based, exact search) & \raggedright Brute-force nearest neighbor search guarantees exact results. & Computationally expensive for large-scale indices; relies on embedding quality. \\
\addlinespace
FAISS - HNSW & \raggedright Dense retrieval (vector-based, approximate search) & \raggedright Allows rapid semantic search through graph-based hops. & Slight reduction in retrieval accuracy due to approximation; relies on embedding quality. \\
\bottomrule
\end{tabularx}
\begin{tablenotes}
\small
\item[a] Node augmentation appended at end of top-k and evaluated with reranker only.
\end{tablenotes}
\end{threeparttable}
\end{table}

\begin{figure}[htbp]
    \centering
    \includegraphics[width=\linewidth]{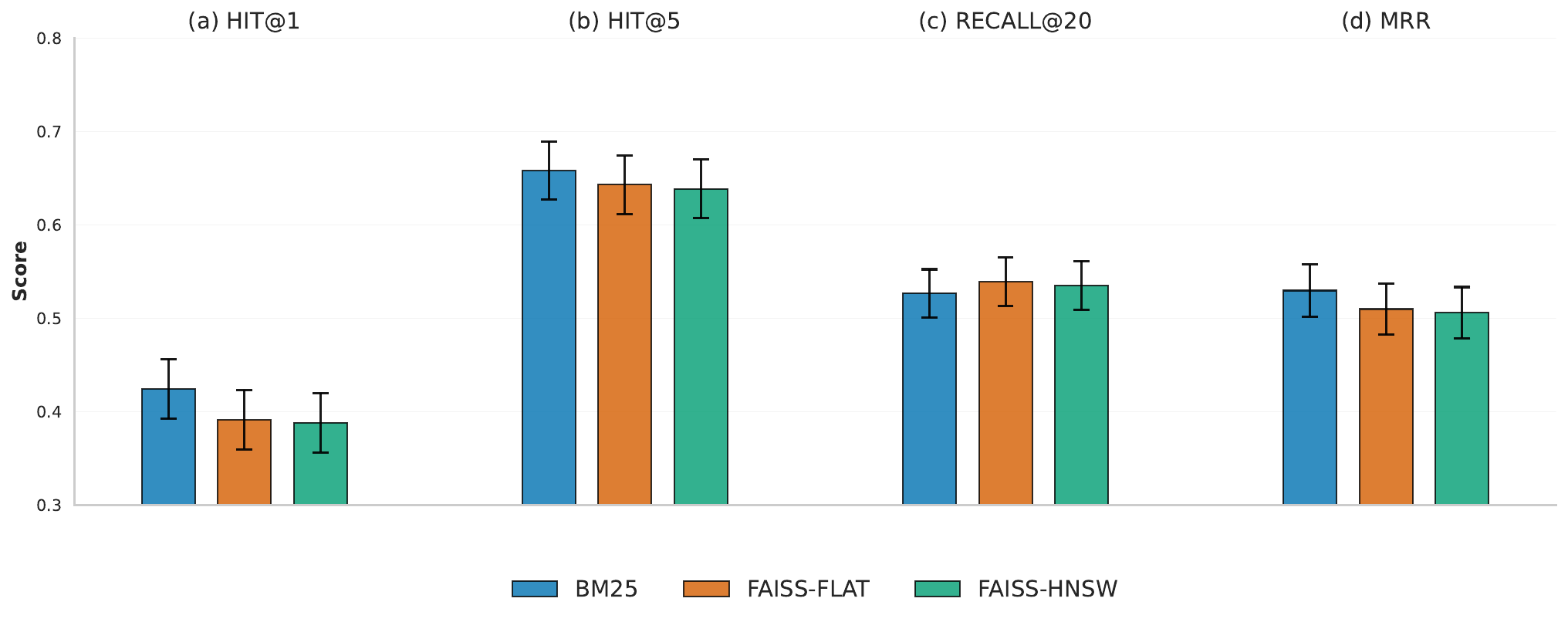}
    \caption{
        \textbf{Performance Comparison of Retrieval Models Across Key Metrics.}
        Four-panel visualization showing performance distributions for three retrieval approaches across \textbf{Hit@1}, \textbf{Hit@5}, \textbf{Recall@20}, and \textbf{MRR} metrics (panels A–D, respectively). Vertical error bars represent 95\% confidence intervals around mean performance scores. Evaluation conducted on validation dataset (n=910 queries).
        }
    \label{fig:four_panel_retriever_ci_plot}
    \Description{
        \textbf{Performance Comparison of Retrieval Models Across Key Metrics.}
        Four-panel visualization showing performance distributions for three retrieval approaches across \textbf{Hit@1}, \textbf{Hit@5}, \textbf{Recall@20}, and \textbf{MRR} metrics (panels A–D, respectively). Vertical error bars represent 95\% confidence intervals around mean performance scores. Evaluation conducted on validation dataset (n=910 queries).
        }
\end{figure}

Table \ref{tab:retriever_results} and Figure \ref{fig:four_panel_retriever_ci_plot} display retriever performance across the four metrics using the Validation query set. This analysis evaluated the performance of three information retrieval pipelines (BM25, FAISS-FLAT, and FAISS-HNSW) across 910 queries using four key metrics: HIT@1, HIT@5, RECALL@20, and MRR. Friedman tests revealed a statistically significant difference between pipelines only for HIT@1 (p = 0.022), where BM25 achieved the highest mean performance (0.424) compared to FAISS-FLAT (0.391) and FAISS-HNSW (0.388), though the effect size was small (Kendall's W = 0.0042). However, post-hoc pairwise comparisons using Bonferroni correction found no significant differences between any pipeline pairs, suggesting that while there may be subtle performance variations, they are not statistically robust when accounting for multiple comparisons. For the remaining metrics—HIT@5, RECALL@20, and MRR—no significant differences were detected between pipelines (all $p > 0.05$), indicating that these three approaches perform similarly across broader retrieval contexts, with BM25 showing modest numerical advantages that do not reach statistical significance.

\begin{table}[htbp]
\caption{Performance Comparison of Retrievers on Validation Dataset (n=910 queries)}
\label{tab:retriever_results}
\centering
\small
\begin{tabular}{@{}lrrrr@{}}
\toprule
\textbf{RAG Model} & \textbf{Hit@1} & \textbf{Hit@5} & \textbf{Recall@20} & \textbf{MRR} \\
\midrule
BM25 & \textbf{0.4242} & \textbf{0.6582} & 0.5264 & \textbf{0.5295} \\
FAISS - FLAT & 0.3912 & 0.6429 & \textbf{0.5390} & 0.5097 \\
FAISS - HNSW & 0.3879 & 0.6385 & 0.5352 & 0.5060 \\
\bottomrule
\end{tabular}
\end{table}


\subsection{Re-ranking Model Comparison}
While numerous cross-encoder and large language models (LLMs) are available for integration into retrieval architectures, this study focused on a systematic comparison of representative models from each category. For LLM-based re-ranking, Claude 4 Sonnet was selected as the primary model due to its recent release, robust API accessibility, and substantial token capacity, which enables processing of longer document contexts during re-ranking. To provide a comprehensive performance comparison, several cross-encoder models detailed in Table \ref{tab:model_overview} were evaluated alongside the LLM approach to assess their relative effectiveness within the proposed retrieval pipeline.

\begin{figure}[htbp]
    \centering
    \includegraphics[width=\linewidth]{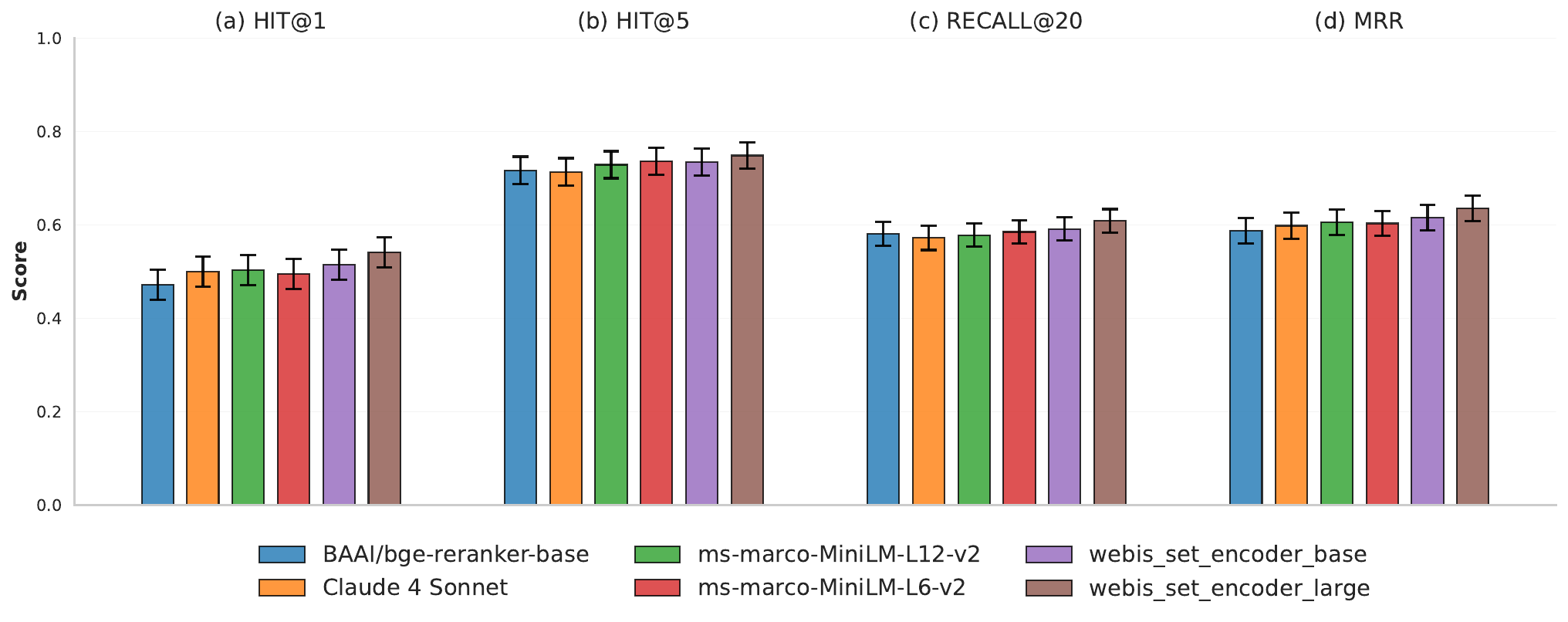}
    \caption{
        \textbf{Performance Comparison of Re-ranking Models Across Key Metrics.}
        Four-panel visualization showing performance distributions for six re-ranking approaches across \textbf{Hit@1}, \textbf{Hit@5}, \textbf{Recall@20}, and \textbf{MRR} metrics (panels A–D, respectively). Vertical error bars represent 95\% confidence intervals around mean performance scores. Evaluation conducted on validation dataset (n=910 queries).
        }
    \label{fig:four_panel_ci_plot}
    \Description{
        \textbf{Performance Comparison of Re-ranking Models Across Key Metrics.}
        Four-panel visualization showing performance distributions for six re-ranking approaches across \textbf{Hit@1}, \textbf{Hit@5}, \textbf{Recall@20}, and \textbf{MRR} metrics (panels A–D, respectively). Vertical error bars represent 95\% confidence intervals around mean performance scores. Evaluation conducted on validation dataset (n=910 queries).
        }
\end{figure}

The evaluation of re-ranking models on the validation dataset (n=910 queries) demonstrates substantial variability in retrieval performance relative to the baseline FAISS HNSW approach without re-ranking. Statistical analysis reveals considerable heterogeneity among re-ranking methods, with top-performing models achieving statistically significant performance gains ($p < 0.001$) across all evaluation metrics, while several models underperformed relative to the baseline retrieval system (Table \ref{tab:ce_reranker_results} and Figure \ref{fig:four_panel_ci_plot}).

The performance improvements were particularly pronounced among the webis set-encoder models and MS Marco cross-encoders, with Hit@1 scores increasing by up to 39.4\% for the best-performing model. The webis/set-encoder-large model achieved the highest Hit@1 performance (0.5407), followed by webis/set-encoder-base (0.5143), ms-marco-MiniL-L-12-v2 (0.5033), and Claude 4 Sonnet (0.5000). Similarly, Hit@5 metrics demonstrated substantial improvements for top-tier models, with webis/set-encoder-large leading this metric (0.7484), representing a 17.3\% improvement over baseline performance. Notably, several models including the NLI-based cross-encoders (nli-deberta-v3-large, nli-roberta-base) and sentence similarity models (stsb-roberta-large) significantly underperformed the baseline, highlighting the importance of model selection in re-ranking architectures.

\begin{table}[htbp]
\caption{Performance Comparison of Rerankers on Validation Dataset (n=910 queries)}
\label{tab:ce_reranker_results}
\small 
\begin{tabularx}{\textwidth}{@{}p{0.50\textwidth}cccc@{}}
\toprule
\textbf{RAG Model} & \textbf{Hit@1} & \textbf{Hit@5} & \textbf{Rec@20} & \textbf{MRR} \\
\midrule
FAISS HNSW + ms-marco-MiniLM-L-6-v2 & 0.4945 & \uline{0.7363}    & 0.5848 & 0.6029 \\
FAISS HNSW + ms-marco-MiniLM-L-12-v2 & 0.5033 & 0.7286 & 0.5779 & 0.6048 \\
FAISS HNSW + amberoad/bert-multilingual & 0.4241 & 0.6890 & 0.5764 & 0.5434 \\
FAISS HNSW + BAAI/bge-reranker-base & 0.4714 & 0.7165 & 0.5806 & 0.5870 \\
FAISS HNSW + webis/set-encoder-base & \uline{0.5143} & 0.7341 & \uline{0.5912} & \uline{0.6147} \\
FAISS HNSW + webis/set-encoder-large & \textbf{0.5407} & \textbf{0.7484} & \textbf{0.6085} & \textbf{0.6355} \\
FAISS HNSW + webis/set-encoder-novelty & 0.3176 & 0.5473 & 0.3942 & 0.4238 \\
FAISS HNSW + nli-deberta-v3-large & 0.0297 & 0.1165 & 0.1401 & 0.0916 \\
FAISS HNSW + stsb-roberta-large & 0.0637 & 0.2165 & 0.2206 & 0.1515 \\
FAISS HNSW + nli-roberta-base & 0.0945 & 0.3143 & 0.1309 & 0.2032 \\
FAISS HNSW + Claude4 Sonnet & 0.5000 & 0.7132 & 0.5716 & 0.5979 \\
\bottomrule
\end{tabularx}
\end{table}

Pairwise statistical analysis using Wilcoxon signed-rank test comparing the six retrieval pipelines across four metrics (Hit@1, Hit@5, Recall@20, MRR) using Bonferroni-corrected significance testing ($\alpha = 0.0033$).
\\
\textbf{Key findings:}
\begin{itemize}
    \item \textbf{Hit@1:} Only 2 significant differences, both favoring webis\_set\_encoder\_large over point-wise models
    \item \textbf{Hit@5:} No significant differences after correction
    \item \textbf{Recall@20:} Most differences detected (6 significant), with set-wise models outperforming others
    \item \textbf{MRR:} 5 significant differences, again favoring webis\_set\_encoder\_large
\end{itemize}

The webis\_set\_encoder\_large consistently outperforms all point-wise cross-encoders across metrics. \textbf{Top performers:} webis\_set\_encoder\_large ($p < 0.0001$), ms-marco-MiniLM-L12-v2 ($p = 0.0006$), and ms-marco-MiniLM-L6-v2 ($p = 0.0006$).


Based on these validation results, the ms-marco-MiniLM-L-6-v2 model was selected for integration into the final pipeline evaluation phase due to its superior Hit@5 and Recall@20 performance and computational efficiency and cost relative to the other model alternatives.

\subsection{Full Retriever-Reranker Pipeline Comparison}\label{sec_pipline}
The final pipeline developed in this study integrates optimized pre-processing, embedding, retrieval, and re-ranking stages to maximize the performance of the RAG system. The best results for each of the various models is in Table \ref{tab:results_summary}. The results for each model utilize optimized embedding practices, hyperparameters, and data-cleaning methods for that model.

\begin{table}[htbp]
\centering
\begin{threeparttable}
\caption{Results Summary -- All Queries}
\label{tab:results_summary}
\small 
\begin{tabularx}{\textwidth}{@{}lcccccccc@{}}
\toprule
& \multicolumn{4}{c}{\textbf{9100 Query Composite}} & \multicolumn{4}{c}{\textbf{6380 Query Composite}} \\
\cmidrule(lr){2-5} \cmidrule(lr){6-9}
\textbf{Model} & \textbf{Hit@1} & \textbf{Hit@5} & \textbf{R@20} & \textbf{MRR} & \textbf{Hit@1} & \textbf{Hit@5} & \textbf{R@20} & \textbf{MRR} \\
\midrule
\rowcolor{gray!10} BM25\tnote{a} & & & & & 0.4494 & 0.6742 & 0.5377 & 0.5530 \\
\rowcolor{gray!10} GritLM-7b\tnote{a} & & & & & 0.4208 & 0.6687 & 0.5652 & 0.5346 \\
\rowcolor{gray!10} ColBERTv2\tnote{a} & & & & & 0.4610 & 0.6602 & 0.5344 & 0.5551 \\
\arrayrulecolor{gray!50}\specialrule{0.6pt}{0pt}{0pt}\arrayrulecolor{black}
BM25 & 0.4337 & 0.6678 & 0.4146 & 0.5404 & 0.4242 & 0.6582 & 0.5264 & 0.5295 \\
BM25+CE\tnote{b} & 0.5182 & 0.7490 & 0.4560 & 0.6211 & 0.5063 & 0.7223 & 0.5770 & 0.6033 \\
BM25+Aug+CE\tnote{b} & 0.5197 & 0.7542 & 0.4649 & 0.6244 & 0.4967 & 0.7374 & 0.5789 & 0.6037 \\
FAISS Flat & 0.4438 & 0.7008 & 0.4320 & 0.5601 & 0.4363 & 0.7088 & 0.4148 & 0.5587 \\
FAISS Flat+CE\tnote{b} & \uline{0.5363} & \uline{0.7803} & \uline{0.4795} & \uline{0.6450} & 0.4989 & \uline{0.7429} & 0.5899 & 0.6081 \\
FAISS HNSW & 0.4413 & 0.6975 & 0.4285 & 0.5573 & 0.3879 & 0.6385 & 0.5351 & 0.5060 \\

FAISS HNSW+CE\tnote{b} & 0.5327 & 0.7756 & 0.4761 & 0.6409 & \uline{0.5125} & 0.7356 & \uline{0.5951} & \uline{0.6128} \\

FAISS HNSW+CE\tnote{c} & \textbf{0.5682} & \textbf{0.7921} & \textbf{0.4929} & \textbf{0.6681} & \textbf{0.5475} & \textbf{0.7525} & \textbf{0.6130} & \textbf{0.6403} \\
\bottomrule
\end{tabularx}
\begin{tablenotes}
\small
\item[a] Benchmark performance metrics on the full dataset as stated in Wu et al.~\cite{wu2024stark}.
\item[b] CE: ms-marco-MiniLM-L-6-v2 Cross-Encoder
\item[c] CE: webis/set-encoder-large Cross-Encoder
\end{tablenotes}
\end{threeparttable}
\end{table}

Statistical analysis using Wilcoxon signed-rank tests with Bonferroni correction ($\alpha = 0.0024$) revealed several key findings across all evaluation metrics. Most importantly, cross-encoder re-ranking provided statistically significant improvements ($p < 0.001$) over baseline retrieval methods across all models and metrics. FAISS-based methods significantly outperformed BM25 approaches ($p < 0.001$), with the webis/set-encoder-large cross-encoder representing the highest-performing configuration. Notably, both FAISS Flat+CE and FAISS HNSW+CE with ms-marco cross-encoders achieved comparable performance, with no statistically significant difference detected in Recall@20.

\begin{figure}[htbp]
    \centering
    \includegraphics[width=1\linewidth]{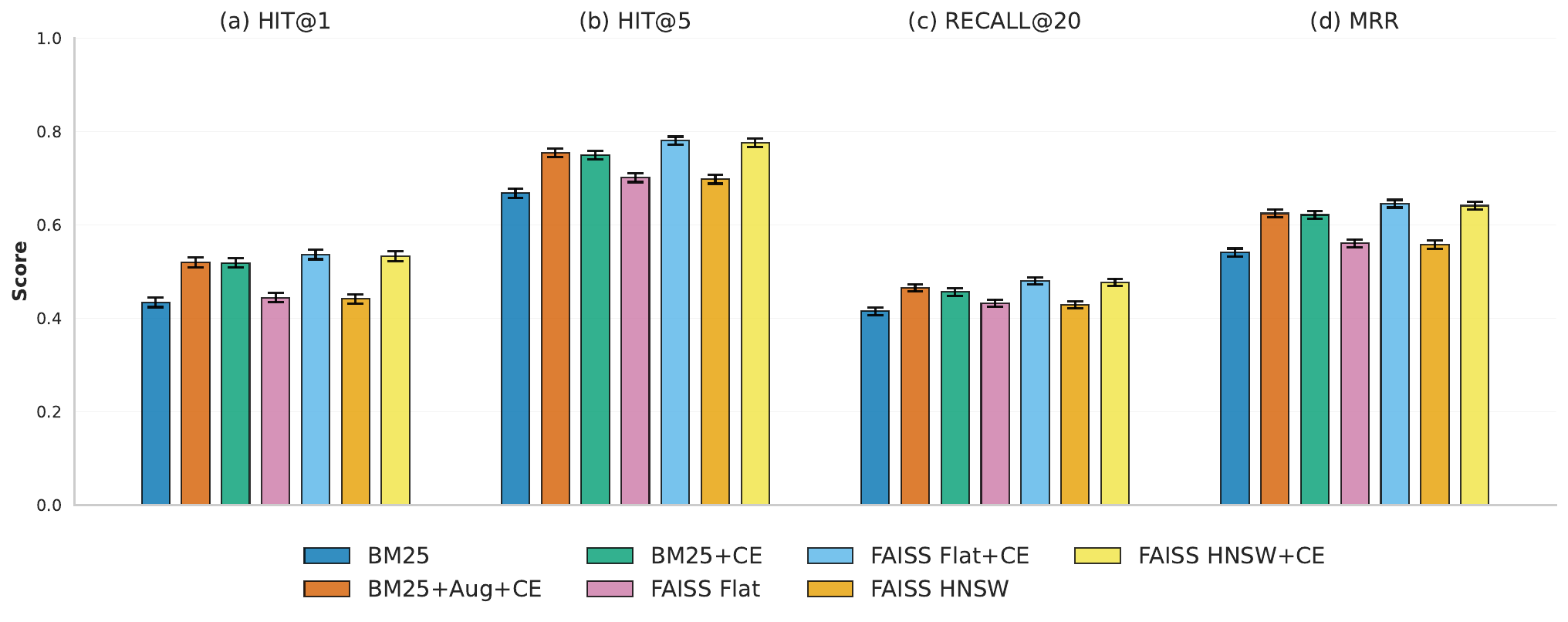}
    \caption{Model Performance Comparison. This four-panel figure presents performance metrics for seven models across \textbf{HIT@1}, \textbf{HIT@5}, \textbf{RECALL@20}, and \textbf{MRR} (panels A–D, respectively). Vertical error bars represent 95\% confidence intervals for each model's mean score. Testing conducted on 9,100 query Evaluation dataset.}
    \Description{Model Performance Comparison. This four-panel figure presents performance metrics for seven models across \textbf{HIT@1}, \textbf{HIT@5}, \textbf{RECALL@20}, and \textbf{MRR} (panels A–D, respectively). Vertical error bars represent 95\% confidence intervals for each model's mean score. Testing conducted on 9,100 query Evaluation dataset.}
    \label{fig:eval_results}
\end{figure}

Figure \ref{fig:eval_results} demonstrates the relative performance of each RAG model. There is significant improvement of the BM25 retrieval with the appending of '1-hop' edge nodes. FAISS models were superior, however, with the FAISS HNSW very closely approximating FAISS Flat's brute force search. While all methods produced good results that improve upon published benchmarks, there is a significant difference in computation time. Due to BM25 running on CPU only and with the appending of additional nodes, the processing time difference is significant. Additionally, the structure of the FAISS HNSW index allows for faster similarity searches compared to FAISS Flat. See Table \ref{tab:time_comparison} for computation times.
All optimized pipeline configurations significantly exceeded published benchmark performance, with our best-performing model (FAISS HNSW + webis/set-encoder-large) achieving Hit@1 improvements of 20.4\% over the strongest published baseline (Claude 3 Reranker: 0.4549 vs. 0.5475). 


However, this superior performance comes with substantial computational costs, as evidenced by the webis/set-encoder-large model's processing speed of 100 seconds per query compared to 0.55 seconds per query for the MS-Marco cross-encoder pipeline (Table \ref{tab:time_comparison}). Balancing performance gains against computational efficiency, we designate FAISS HNSW + MS-Marco cross-encoder as the pipeline of choice for practical deployment scenarios, while acknowledging that FAISS HNSW + webis/set-encoder-large represents the performance ceiling for applications where computational resources are not constrained.

\begin{table}[htbp]
\begin{threeparttable}
\caption{Computation Time Comparison (9,100 Queries)}
\label{tab:time_comparison}
\small
\begin{tabularx}{\textwidth}{@{}p{6.5cm}>{\centering\arraybackslash}p{2.0cm}>{\centering\arraybackslash}p{2.0cm}>{\centering\arraybackslash}p{2.0cm}@{}}
\toprule
\textbf{Model} & \makecell{\textbf{Retriever}\\(sec/query)} & \makecell{\textbf{Reranker}\\(sec/query)} & \makecell{\textbf{Total}\\(sec/query)} \\
\midrule
BM25 + Augment + MS Marco cross-encoder & 7.69 & 2.27 & 9.96 \\
FAISS Flat + MS Marco cross-encoder & 0.42 & 0.55 & 0.97 \\
FAISS HNSW + MS Marco cross-encoder & 0.02 & 0.53 & 0.55 \\
FAISS HNSW + Webis set-encoder-large & 0.02 & 100 & 100.02 \\
\bottomrule
\end{tabularx}
\begin{tablenotes}
\small
\item Note: All performance benchmarks conducted in Google Colab environment with Tesla T4 GPU.
\end{tablenotes}
\end{threeparttable}
\end{table}


\section{Conclusion}\label{sec5}

This study demonstrates that strategically combining FAISS-based dense retrieval with cross-encoder re-ranking achieves substantial performance improvements for knowledge extraction from semi-structured databases. Our systematic evaluation on the Amazon STaRK dataset reveals that cross-encoder re-ranking consistently and significantly enhances retrieval performance across all configurations. The three key findings from this study are: (1) FAISS-based dense retrieval consistently outperforms BM25 sparse retrieval, even with graph augmentation; (2) HNSW indexing achieves performance comparable to brute-force search with significant speed advantages; and (3) specialized cross-encoders outperform general-purpose LLMs for re-ranking tasks.

These results have immediate practical implications for e-Commerce platforms and similar domains managing large-scale proprietary knowledge in SKB format. The significant difference in processing speed between the faster FRMR and more accurate FRWSR provides an indication of which architectures are more fitting for different applications. For example, FRMR is likely better suited for customer-facing applications requiring fast response times, while back-end analytics systems may prioritize the accuracy of FRWSR. The success of dense retrieval over sparse methods also suggests that the semantic embedding infrastructure is more powerful than traditional keyword-based systems. However, future research may consider applying similar graph augmentation methods used in conjunction with sparse embedding in this study to other architectures, to better leverage the information in graph databases for retrieval.

This study also affirms cross-encoders as more efficient and accurate alternatives to LLM-based reranker components. The superior performance of specialized cross-encoders, particularly the Webis set-encoder's inter-passage attention mechanism \cite{schlatt_2025}, demonstrates that architectural innovations targeted at retrieval tasks can outperform general-purpose language models while requiring substantially fewer computational resources. This finding aligns with recent work suggesting that task-specific architectures may be more valuable than scale alone \cite{dejean2024reranking}. Set-Encoder is particularly effective when considering inter-passage novel information, making it superior to point-wise models, which allows RAG to avoid redundant information.

Our study is limited by its focus on a single e-Commerce SKB and a synthetic/curated query set; compute constraints also shaped some choices, and we did not fully exploit graph structure during generation. These boundaries suggest clear avenues for further work. For example, graph-aware retrievers and rerankers that natively model edges and multi-hop paths could better exploit relational information \cite{peng2024graphrag}. Additionally, hybrids that fuse sparse and dense signals with query-adaptive top-k and orchestrate early exit reranking may retain quality while reducing cost \cite{luan2021sparse}. Ultimately, future work should investigate the scalability of larger knowledge graphs and domain adaptation techniques to enhance their broader applicability.

Overall, the showcased Retriever-Reranker architectures establish a new performance baseline for the Amazon STaRK dataset and offers a scalable architecture for similar knowledge-grounded retrieval applications. With these results, this study demonstrates that modular, open-source components can achieve competitive performance in enterprise-scale information retrieval, providing practitioners with empirically validated approaches for optimizing retrieval in semi-structured knowledge environments.

\bibliographystyle{ACM-Reference-Format}
\bibliography{references}





\end{document}